\documentclass[pra,aps,showpacs,twocolumn]{revtex4}

\usepackage{amsmath}
\usepackage{bm}
\usepackage{graphicx}

\begin{document}

\title{Capillary instability in a two-component Bose--Einstein condensate}

\author{Kazuki Sasaki}
\author{Naoya Suzuki}
\author{Hiroki Saito}

\affiliation{
Department of Engineering Science, University of Electro-Communications,
Tokyo 182-8585, Japan
}

\date{\today}

\begin{abstract}
Capillary instability and the resulting dynamics in an immiscible
two-component Bose--Einstein condensate are investigated using the
mean-field and Bogoliubov analyses.
A long, cylindrical condensate surrounded by the other component is
dynamically unstable against breakup into droplets due to the
interfacial tension arising from the quantum pressure and interactions.
A heteronuclear system confined in a cigar-shaped trap is proposed for
realizing this phenomenon experimentally.
\end{abstract}


\maketitle

\section{Introduction}

A fluid cylinder will be unstable against breakup into droplets if its
length exceeds its circumference.
A familiar example is a water jet projected into air, which breaks up into
many water droplets.
Savart~\cite{Savart} first investigated this kind of instability and 
Plateau~\cite{Plateau} subsequently developed an experimental technique.
Lord Rayleigh~\cite{Rayleigh} used linear stability analysis to
investigate this phenomenon.
This instability is known as capillary instability~\cite{Chandra}
(or sometimes as Plateau--Rayleigh instability).
The present paper proposes a method for observing capillary instability
in quantum degenerate gases.

Classical fluid instabilities have recently been studied theoretically in
quantum fluids with renewed interest.
Various fluid instabilities are predicted to be observable in
two-component Bose--Einstein condensates (BECs) including the
Rayleigh--Taylor instability~\cite{Sasaki,Gautam}, the Kelvin--Helmholtz
instability~\cite{Takeuchi,Suzuki}, the Richtmyer--Meshkov
instability~\cite{Bezett}, the counter-superflow
instability~\cite{Law,TakeuchiL}, and the Rosensweig
instability~\cite{SaitoL}.
Of these, the counter-superflow instability has recently been realized
experimentally~\cite{Hoefer}.
The B\'enard--von K\'arm\'an vortex street, which is predicted to occur
in single-component~\cite{SasakiL} and two-component BECs~\cite{Sasaki2},
is also caused by fluid instability.

In the present paper, we investigate capillary instability and the
resulting dynamics in a two-component BEC.
Capillary instability in a water jet originates from the surface tension
of water that results from the attractive interaction between water
molecules.
By contrast, in the present study, we demonstrate that capillary
instability emerges in an atomic gas with repulsive inter- and
intra-atomic interactions.
We consider an immiscible two-component BEC in which a cylinder of
component 1 is surrounded by component 2.
The interfacial tension between the two components~\cite{Barankov,Schae}
induces capillary instability and component 1 breaks into droplets that
form a matter-wave soliton train.
This system thus exhibits capillary instability and is novel in that the
instability is only due to repulsive interactions.
We also demonstrate that dark--bright soliton and skyrmion trains can be
generated in this system.

The present paper is organized as follows.
Section~\ref{s:ideal} numerically demonstrates capillary instability and
subsequent dynamics in an ideal system.
Section~\ref{s:bogo} examines the stability of the system using Bogoliubov
and linear stability analyses.
Section~\ref{s:trap} proposes a realistic system for observing capillary
instability in a trapped two-component BEC and numerically demonstrates
various dynamics.
Section~\ref{s:conc} presents the conclusions of this study.

\section{Dynamics in an ideal system}
\label{s:ideal}

We consider a two-component BEC of atoms with mass $m_j$ in an external
potential $V_j$, where the subscript $j = 1$, $2$ indicates component 1 or
2.
In zero-temperature mean-field theory, the macroscopic wave functions
$\psi_1$ and $\psi_2$ obey the Gross--Pitaevskii (GP) equation given by
\begin{subequations} \label{GP}
\begin{eqnarray}
i \hbar \frac{\partial \psi_1}{\partial t} & = & \left(
-\frac{\hbar^2}{2m_1} \nabla^2 + V_1 + g_{11} |\psi_1|^2 + g_{12}
|\psi_2|^2 \right) \psi_1,
\nonumber \\
\\
i \hbar \frac{\partial \psi_2}{\partial t} & = & \left(
-\frac{\hbar^2}{2m_2} \nabla^2 + V_2 + g_{22} |\psi_2|^2 + g_{12}
|\psi_1|^2 \right) \psi_2,
\nonumber \\
\end{eqnarray}
\end{subequations}
where the interaction parameters are defined as
\begin{equation}
g_{jj'} = 2\pi \hbar^2 a_{jj'} \left( m_j^{-1} + m_{j'}^{-1} \right),
\end{equation}
with $a_{jj'}$ being the $s$-wave scattering length between the atoms in
components $j$ and $j'$ $(j, j' = 1, 2)$.
We assume that the interaction parameters satisfy
\begin{equation} \label{sepa}
g_{11} g_{22} < g_{12}^2,
\end{equation}
for which the two components are immiscible~\cite{Pethick}.

In this section, we consider an ideal system with $V_1 = V_2 = 0$, $m_1
= m_2 \equiv m$, and $g_{11} = g_{22} \equiv g$.
Normalizing the length and time in Eq.~(\ref{GP}) by $\xi = \hbar / (m g
n)^{1/2}$ and $\xi / v_{\rm s}$, where $n$ is the atomic density
$|\psi_2|^2$ far from the interface and $v_{\rm s} = (g n / m)^{1/2}$ is
the sound velocity, we find that the intracomponent interaction
coefficients are normalized to unity and the relevant parameter is only
$g_{12} / g$.
The initial state is a stationary state $\Psi_j(r)$ in which component 1
is a cylinder centered about the $z$ axis and is surrounded by component
2.
The state $\Psi_j(r)$ depends only on $r = (x^2 + y^2)^{1/2}$ and has
cylindrical symmetry and translation symmetry along $z$.
Numerically, this initial state is prepared by imaginary time propagation
of the GP equation (\ref{GP}).
The time evolution of the system is obtained by numerically solving the GP
equation (\ref{GP}) by the pseudospectral method.

\begin{figure}[t]
\includegraphics[width=8.5cm]{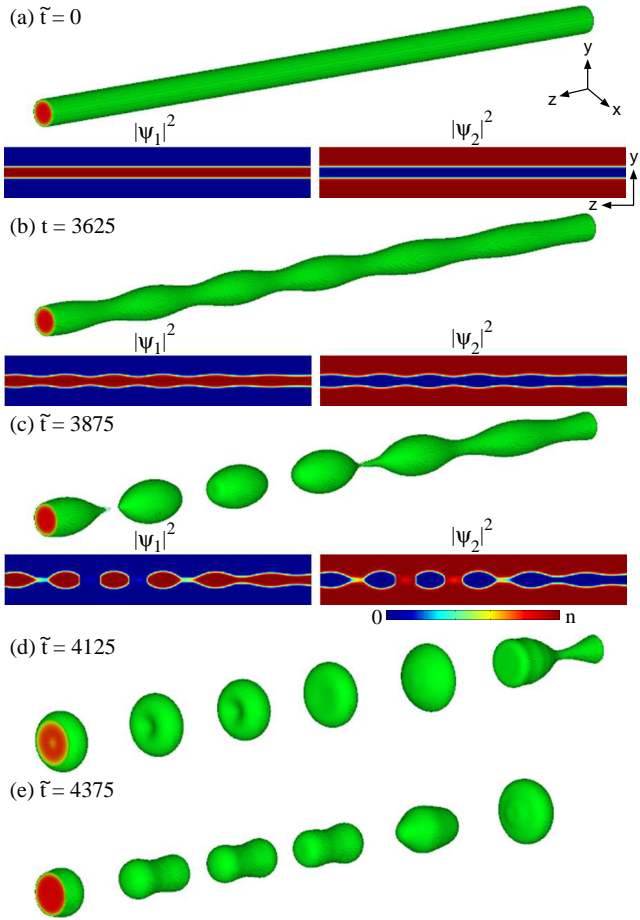}
\caption{
(Color online) Capillary instability and resulting dynamics of a
two-component BEC for $g_{12} / g = 1.1$.
The three-dimensional objects depict the isodensity surface of component
1.
The lower panels in (a)--(c) show the density distributions $|\psi_j|^2$
at the cross section of $x = 0$ (the field of view is $600 \xi \times
100 \xi$).
The number of component 1 atoms in an initial cylinder with length $\xi$
is $N_{1\xi} = 400$.
}
\label{f:ideal}
\end{figure}
Figure~\ref{f:ideal} shows the time evolution of the system for $g_{12} /
g = 1.1$, where the number of component 1 atoms in an initial cylinder with
length $\xi$,
\begin{equation}
N_{1\xi} \equiv \xi \int dx dy |\Psi_1|^2,
\end{equation}
is 400.
We add a small white noise to the initial state to break the exact
numerical symmetry.
At $\tilde t \equiv t v_{\rm s} / \xi = 3600$, ``varicose''~\cite{Chandra}
modulation with a wavelength $\simeq 99 \xi \simeq 9.0 R_0$ occurs on the
cylindrical interface [Fig.~\ref{f:ideal} (b)], where $R_0$ is the initial
radius of the cylinder of component 1.
The varicosities then split into droplets [Fig.~\ref{f:ideal} (c)], which
undergo quadrupole oscillations [Figs.~\ref{f:ideal} (d) and \ref{f:ideal}
(e)] due to the kinetic energy converted from the interfacial tension
energy.
The droplets subsequently aggregate to minimize the interfacial tension
energy.

\begin{figure}[t]
\includegraphics[width=8.5cm]{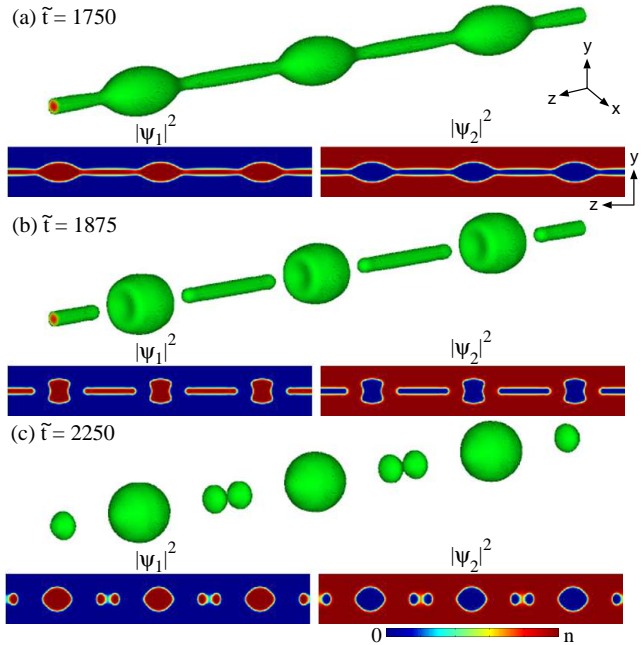}
\caption{
(Color online) Time evolution of the system with a periodic initial
perturbation given by Eq.~(\ref{periodic}).
The parameters are the same as those in Fig.~\ref{f:ideal}.
The three-dimensional objects indicate the isodensity surfaces of
component 1 and the lower panels show the density distributions
$|\psi_j|^2$ at the cross section of $x = 0$ (the field of view is $600
\xi \times 100 \xi$).
}
\label{f:sate}
\end{figure}
Figure~\ref{f:sate} shows the time evolution of the system, where a
periodic initial perturbation is added to component 1 as
\begin{equation} \label{periodic}
\Psi_1(r) \left( 1 + 0.01 \cos \frac{2 \pi z}{200 \xi} \right),
\end{equation}
instead of the white noise.
This wavelength $200 \xi$ is about twice the most unstable wavelength
in the modulation in Fig.~\ref{f:ideal} (b).
The sinusoidal seed develops into a highly nonlinear pattern
[Fig.~\ref{f:sate} (a)] and the thin ligaments separate from the main
droplets [Fig.~\ref{f:sate} (b)], which deform into small drops
[Fig.~\ref{f:sate} (c)].
Such nonlinear behavior is very similar to that in classical liquid
jets~\cite{Bogy}, and the small drops are termed ``satellite drops''.

\section{Stability analysis}
\label{s:bogo}

\subsection{Bogoliubov analysis}

\begin{figure}[t]
\includegraphics[width=8.5cm]{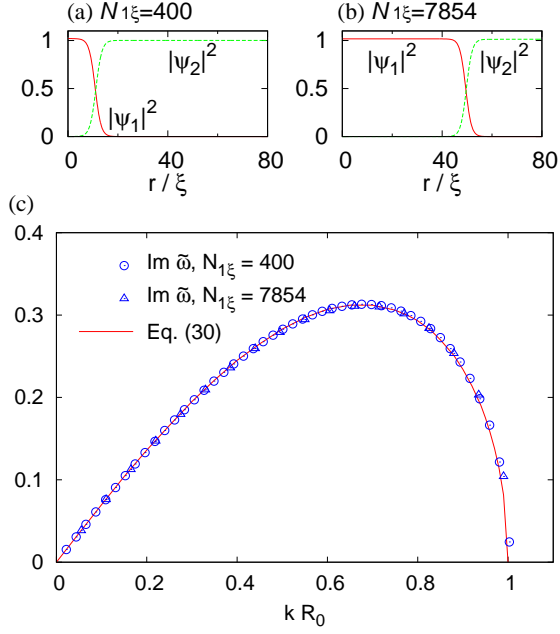}
\caption{
(Color online) Radial density profiles of the stationary states for (a)
	$N_{1\xi} = 400$ and (b) $N_{1\xi} = 7854$ with $g_{12} / g = 1.1$.
(c) Imaginary part of the excitation frequencies.
The circles and triangles are obtained by numerical diagonalization of
Eq.~(\ref{BdG}) with the parameters used in (a) and (b), respectively.
The plotted quantities are normalized using Eq.~(\ref{nomega}) with $R_0 =
11 \xi$ for the circles and $R_0 = 50 \xi$ for the triangles.
The solid curve indicates Eq.~(\ref{analytic}).
}
\label{f:bogo}
\end{figure}
We study the stability of an ideal system such as that depicted in
Fig.~\ref{f:ideal} (a) using Bogoliubov theory.
We separate the wave function as
\begin{equation} \label{delpsi}
\psi_j(r, z, t) = \left[ \Psi_j(r) + \delta\psi_j(r, z, t) \right] e^{-i
\mu_j t / \hbar},
\end{equation}
where $\mu_j$ is the chemical potential and the system is assumed to have
cylindrical symmetry.
Figures~\ref{f:bogo} (a) and \ref{f:bogo} (b) show examples of the
stationary state $\Psi_j(r)$.
Substituting Eq.~(\ref{delpsi}) into the GP equation (\ref{GP}) and taking
the first order of the small deviation $\delta \psi_j$, we find $(j \neq
j')$
\begin{eqnarray} \label{delpsi2}
i \hbar \frac{\partial \delta \psi_j}{\partial t} & = & \left(
-\frac{\hbar^2}{2m_j} \nabla^2 + V_j - \mu_j + 2 g_{jj} \Psi_j^2 +
g_{jj'} \Psi_{j'}^2 \right) \delta \psi_j 
\nonumber \\
& & + g_{jj} \Psi_j^2 \delta\psi_j^* + g_{jj'} \Psi_j \Psi_{j'}
(\delta\psi_{j'} + \delta\psi_{j'}^*),
\end{eqnarray}
where $\Psi_1$ and $\Psi_2$ are assumed to be real without loss of
generality.
Using the excitation mode of the form
\begin{equation}
\delta\psi_j = u_{jk}(r) e^{i (kz-\omega t)} + v_{jk}^*(r)
e^{-i (kz-\omega t)}
\end{equation}
in Eq.~(\ref{delpsi2}), we obtain the Bogoliubov--de Gennes equation $(j 
\neq j')$,
\begin{widetext}
\begin{subequations} \label{BdG}
\begin{eqnarray}
\left[ -\frac{\hbar^2}{2m_j} \left( \frac{\partial^2}{\partial r^2} +
	\frac{1}{r} \frac{\partial}{\partial r} - k^2 \right) + V_j - \mu_j +
	2 g_{jj} \Psi_j^2 + g_{jj'} \Psi_{j'}^2 \right] u_{jk} + g_{jj} \Psi_j^2
	v_{jk}
+ g_{jj'} \Psi_j \Psi_{j'} (u_{j'k} + v_{j'k}) & = & \hbar \omega
u_{jk}, \\
\left[ -\frac{\hbar^2}{2m_j} \left( \frac{\partial^2}{\partial r^2} +
	\frac{1}{r} \frac{\partial}{\partial r} - k^2 \right) + V_j - \mu_j +
	2 g_{jj} \Psi_j^2 + g_{jj'} \Psi_{j'}^2 \right] v_{jk} + g_{jj} \Psi_j^2
	u_{jk}
+ g_{jj'} \Psi_j \Psi_{j'} (u_{j'k} + v_{j'k}) & = & -\hbar \omega
	v_{jk}.
\end{eqnarray}
\end{subequations}
\end{widetext}

Figure~\ref{f:bogo} (c) shows the imaginary part of the Bogoliubov
excitation frequency obtained by numerical diagonalization of
Eq.~(\ref{BdG}).
For comparison with an analytic result given in the next subsection, we
plot the normalized quantity as
\begin{equation} \label{nomega}
{\rm Im} \tilde \omega \equiv \sqrt{\frac{m n R_0^3}{\alpha}} {\rm Im}
\omega,
\end{equation}
where $\alpha$ is the interfacial tension coefficient originating from the
excess kinetic and interaction energies at the interface.
For $g_{12} / g = 1.1$, we can use the expression for the interfacial
tension coefficient obtained in Refs.~\cite{Barankov,Schae}, namely,
\begin{equation}
\alpha = \frac{\hbar n^{3/2}}{\sqrt{2m}} \sqrt{g_{12} - g}.
\end{equation}
Figure~\ref{f:bogo} (c) reveals that the most unstable wavelength is
$\simeq 9.3 R_0$, which agrees with the modulation in Fig.~\ref{f:ideal}
(b).
In Fig.~\ref{f:bogo} (c), we plot two cases for the parameters
corresponding to Fig.~\ref{f:bogo} (a) (circles) and Fig.~\ref{f:bogo} (b)
(triangles).
We note that these plots fall on a universal curve using the scaling in
Eq.~(\ref{nomega}).

\subsection{Linear stability analysis}

To understand the Bogoliubov spectrum obtained in Fig.~\ref{f:bogo}, we
analyze the stability of the system using a similar method to Rayleigh's
linear stability analysis~\cite{Rayleigh,Chandra}.
We start from the mean-field Lagrangian given by
\begin{equation} \label{L}
L = \int d\bm{r} \left( P_1 + P_2 - g_{12} |\psi_1|^2 |\psi_2|^2 \right),
\end{equation}
where
\begin{equation} \label{P}
P_j = i \hbar \psi_j^* \frac{\partial \psi_j}{\partial t} -
\frac{\hbar^2}{2m_j} |\bm{\nabla} \psi_j|^2 - V_j |\psi_j|^2 -
\frac{g_{jj}}{2} |\psi_j|^4.
\end{equation}
The functional derivative of the action $\int L dt$ with respect to
$\psi_j^*$ gives the GP equation (\ref{GP}).
In this subsection, we assume that the two components are strongly phase
separated and that the interface thickness is negligibly small.
We also assume that the system has cylindrical symmetry and that the
interface is located at $r = R(z, t)$.
We then approximate Eq.~(\ref{L}) as
\begin{equation} \label{L2}
L = 2\pi \int dz \left( \int_0^{R} r dr P_1 + \int_R^\infty r dr P_2
\right) - \alpha S,
\end{equation}
where $S$ is the area of the interface.
Differentiating Eq~(\ref{L2}) with respect to $R$, we obtain~\cite{Landau}
\begin{equation} \label{Lap}
P_1(R, z, t) - P_2(R, z, t) = \alpha \left( \frac{1}{R_1} + \frac{1}{R_2}
\right),
\end{equation}
where $R_1$ and $R_2$ are the principal radii of the interface curvature.
Equation (\ref{Lap}) corresponds to Laplace's formula in fluid mechanics.

We write the wave function as
\begin{equation}
\psi_j(r, z, t) = \sqrt{n_j(r, z, t)} e^{i \phi_j(r, z, t)},
\end{equation}
where $n_j$ and $\phi_j$ are real functions, $n_1 = 0$ for $r > R$, and
$n_2 = 0$ for $r < R$.
We separate the density and phase as
\begin{subequations} \label{sep}
\begin{eqnarray}
n_j(r, z, t) & = & n_{j0} + \delta n_j(r, z, t), \\
\phi_j(r, z, t) & = & -\frac{g_{jj} n_{j0}}{\hbar} t + \delta \phi_j(r, z,
t),
\end{eqnarray}
\end{subequations}
and substitute them into Eq.~(\ref{GP}).
Taking the first order of $\delta \phi_j$ and $g_{jj} \delta n_j$, we have
\begin{eqnarray}
\label{divv}
\bm{\nabla} \cdot \bm{v}_j & = & 0, \\
\hbar \frac{\partial \delta \phi_j}{\partial t} + g_{jj} \delta n_j & = &
0, \label{Ber}
\end{eqnarray}
where the velocity is defined as
\begin{equation}
\bm{v}_j = \frac{\hbar}{m_j} \bm{\nabla} \delta \phi_j.
\end{equation}
The gradient of Eq.~(\ref{Ber}) gives
\begin{equation} \label{Ber2}
m_j \frac{\partial \bm{v}_j}{\partial t} + g_{jj} \bm{\nabla} \delta n_j =
0.
\end{equation}
Substituting Eq.~(\ref{sep}) into Eq.~(\ref{P}) and using Eq.~(\ref{Ber}),
we obtain the pressure,
\begin{equation} \label{press}
P_j = \frac{1}{2} g_{jj} n_{j0}^2 + g_{jj} n_{j0} \delta n_j.
\end{equation}

We assume the sinusoidal forms of the small excitation as
\begin{subequations} \label{small}
\begin{eqnarray}
R(z, t) & = & R_0 + \epsilon \sin(kz - \omega t), \label{R} \\
v_{jr}(r, z, t) & = & \tilde{v}_{jr}(r) \cos(kz - \omega t), \label{vjr}
\\
v_{jz}(r, z, t) & = & \tilde{v}_{jz}(r) \sin(kz - \omega t), \\
\delta n_j & = & \delta \tilde{n}_j(r) \sin(kz - \omega t). \label{dnj}
\end{eqnarray}
\end{subequations}
Using Eqs.~(\ref{press}) and (\ref{small}), Eq.~(\ref{Lap}) becomes
\begin{equation} \label{Lap2}
g_{11} n_{10} \delta \tilde n_1(r) - g_{22} n_{20} \delta \tilde n_2(r)
= \frac{\alpha \epsilon}{R_0^2} (k^2 R_0^2 - 1),
\end{equation}
where we have used the equilibrium condition $(g_{11} n_{10}^2 - g_{22}
n_{20}^2) / 2 = \alpha / R_0$.
Substituting Eqs.~(\ref{vjr})--(\ref{dnj}) into Eqs.~(\ref{divv}) and
(\ref{Ber2}) yields
\begin{equation}
r^2 \tilde{v}_{jr}''(r) + r \tilde{v}_{jr}'(r) - (1 + k^2 r^2)
\tilde{v}_{jr}(r) = 0.
\end{equation}
The solutions of this differential equation are
\begin{subequations} \label{bes}
\begin{eqnarray}
\tilde{v}_{1r}(r) & = & c_1 I_1(kr), \\
\tilde{v}_{2r}(r) & = & c_2 K_1(kr),
\end{eqnarray}
\end{subequations}
where $c_1$ and $c_2$ are constants and $I_n$ and $K_n$ are modified
Bessel functions of the first and second kinds.
These functions are chosen such that components 1 and 2 do not diverge at
the $z$ axis and infinity, respectively.
It follows from Eqs.~(\ref{Ber2}) and (\ref{bes}) that
\begin{subequations} \label{bes2}
\begin{eqnarray}
g_{11} \delta \tilde{n}_1(r) & = & -\frac{m_1 \omega c_1}{k} I_0(kr), \\
g_{22} \delta \tilde{n}_2(r) & = & \frac{m_2 \omega c_2}{k} K_0(kr).
\end{eqnarray}
\end{subequations}
The kinematic boundary condition at the interface is $dR / dt = v_{1r}(R)
= v_{2r}(R)$, which gives
\begin{equation} \label{kin}
\epsilon \omega = -c_1 I_1(R_0) = -c_2 K_1(R_0).
\end{equation}

Using Eqs.~(\ref{Lap2}), (\ref{bes2}), and (\ref{kin}), we
obtain the dispersion relation of the excitation as
\begin{equation} \label{disp}
\omega^2 = \frac{\alpha}{R_0^3} \frac{k R_0 (k^2 R_0^2 - 1)}{m_1 n_{10}
\frac{I_0(k R_0)}{I_1(k R_0)} + m_2 n_{20} \frac{K_0(k R_0)}{K_1(k
R_0)}}.
\end{equation}
This dispersion relation has the same form as that of classical inviscid
incompressible fluids~\cite{Chris}.
The quantum mechanical pressure is included in the interfacial tension
coefficient $\alpha$ and Eq.~(\ref{disp}) does not contain any explicit
quantum correction term.
For $0 < k R_0 < 1$, the right-hand side of Eq.~(\ref{disp}) is negative
and the frequency $\omega$ is pure imaginary.
The mode with a wavelength larger than $2 \pi R_0$ is therefore
dynamically unstable.
The most unstable wave number is given by $k R_0 \simeq 0.68$ and the most
unstable wavelength is $\simeq 9.2 R_0$, which is in good agreement with
the modulation wavelength in Fig.~\ref{f:ideal} (b).
The most unstable wave number depends only on $R_0$ and is independent of
the interaction parameters.
The interaction, which is included in $\alpha$, only affects the growth
rate of the unstable modes.

The solid line in Fig.~\ref{f:bogo} indicates
\begin{equation} \label{analytic}
\sqrt{\frac{m n R_0^3}{\alpha}} {\rm Im} \omega
 = \sqrt{\frac{k R_0 (1 - k^2 R_0^2)}{\frac{I_0(k R_0)}{I_1(k
	R_0)} + \frac{K_0(k R_0)}{K_1(k R_0)}}}
\end{equation}
for $0 < k R_0 < 1$, which is in excellent agreement with the numerically
obtained Bogoliubov spectrum normalized by Eq.~(\ref{nomega}).
This confirms that the modulation instability shown in Fig.~\ref{f:ideal}
(b) is the capillary instability.
Figure~\ref{f:bogo} shows that the dispersion relation (\ref{disp}) is
accurate even when the radius of the cylinder of component 1 is of the
same order as the interface thickness.

\section{Dynamics in a trapped system}
\label{s:trap}

We propose a realistic experimental situation to observe the capillary
instability in a trapped BEC.
We consider a heteronuclear condensate consisting of $^{41}{\rm K}$
(component 1) and $^{87}{\rm Rb}$ (component 2) atoms in the stretched $|F
= 2, m_F = 2 \rangle$ states.~\cite{Modugno}.
The scattering lengths are $a_{11} = 65 a_{\rm B}$~\cite{Wang}, $a_{22} =
99 a_{\rm B}$~\cite{Marte}, and $a_{12} = 163 a_{\rm B}$~\cite{Ferlaino},
which satisfy the phase separation condition in Eq.~(\ref{sepa}).
The two components are confined in axisymmetric harmonic potentials,
\begin{equation}
V_j = \frac{m_j}{2} (\omega_{j\perp}^2 r^2 + \omega_{j z}^2 z^2),
\end{equation}
produced by laser beams.
Since the electronic excitation frequencies of the two atomic species are
separated from each other, $V_1$ and $V_2$ can be chosen independently by
using laser beams with different frequencies.

The initial state is prepared for trap frequencies satisfying $\omega_{1
\perp} \gg \omega_{2 \perp}$ and $\omega_{j \perp} \gg \omega_{j z}$.
Since the radial confinement of component 1 is much tighter than that of
component 2, the ground state of component 1 is extremely narrow and it is
surrounded by component 2 in the radial direction.
The tight radial confinement prevents component 1 from breaking up into
droplets for the same reason as why a classical liquid in a pipe never
exhibits the capillary instability.
We then add small white noise to the initial state and reduce $\omega_{1
\perp}$ suddenly at $t = 0$. 
This produces an unstable initial state similar to the state in
Fig.~\ref{f:ideal} (a), and capillary instability is obtained.

\begin{figure}[t]
\includegraphics[width=8.5cm]{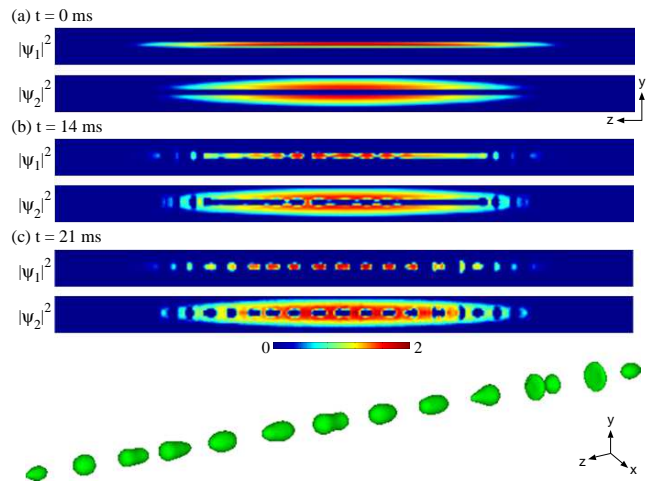}
\caption{
(Color online) Time evolution of a $^{41}{\rm K}$-$^{87}{\rm Rb}$
condensate confined in a cigar-shaped trap.
The initial state is the ground state, in which $N_1 = 5 \times 10^4$
$^{41}{\rm K}$ atoms (component 1) are confined in $V_1$ with $(\omega_{1
\perp}, \omega_{1 z}) / 2\pi = (582, 11.6)$ Hz and $N_2 = 9.5 \times 10^5$
$^{87}{\rm Rb}$ atoms (component 2) are confined in $V_2$ with $(\omega_{2
\perp}, \omega_{2 z}) / 2\pi = (89.4, 7.2)$ Hz.
At $t = 0$, $\omega_{1 \perp} / (2\pi)$ is changed from $582$ Hz to $145$
Hz.
The panels in (a)--(c) show the density distributions $|\psi_j|^2$ at the
cross section of $x = 0$, where the unit of the density is $10^{20}$ ${\rm
m}^{-3}$ and the field of view is $220 \times 14$ $\mu{\rm m}$.
The bottom figure shows the isodensity surface of component 1 at $t = 21$
ms.
}
\label{f:trap}
\end{figure}
Figure~\ref{f:trap} demonstrates the capillary instability and resulting
dynamics of a trapped two-component BEC obtained by numerically solving
the 3D GP equation.
The ratio of the number of atoms in component 1 to the total number of
atoms is $N_1 / (N_1 + N_2) = 0.05$.
At $t = 0$, the radial confinement of component 1 is relaxed and
modulation grows due to the capillary instability.
Varicose modulation occurs in component 1 [Fig.~\ref{f:trap} (b)], which
breaks up into droplets [Fig.~\ref{f:trap} (c)].
The axisymmetry of the system is preserved within this time scale.
We have thus shown that the capillary instability can be observed in a
trapped two-component BEC.
The structure in Fig.~\ref{f:trap} (c) is similar to a matter-wave soliton
train.
A matter-wave soliton and soliton train have so far been produced by a
single-component BEC with attractive interactions~\cite{Khay,Strecker}.
By contrast, the train of matter-wave droplets in Fig.~\ref{f:trap} (c) is
produced only by repulsive interactions.

\begin{figure}[t]
\includegraphics[width=8.5cm]{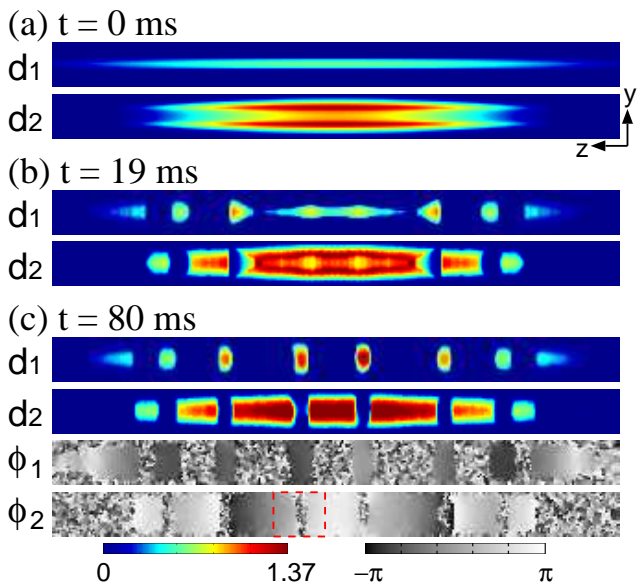}
\caption{
(Color online) Time evolution for $N_1 = 2 \times 10^5$ and $N_2 = 8
\times 10^5$.
The trap frequencies are $(\omega_{1\perp}, \omega_{1z}) / 2\pi = (436,
35)$ and $(\omega_{2\perp}, \omega_{2z}) / 2\pi = (89.4, 7.2)$.
At $t = 0$, $\omega_{1\perp} / 2\pi$ is changed to 145 Hz.
The panels show the column density $d_j = \int |\psi_j|^2 dx$ and the
phase $\phi_j = {\rm arg} \psi_j(x = 0)$.
The unit of the column density is $10^{15}$ ${\rm m}^{-2}$ and the field
of view is $170 \times 13.6$ $\mu{\rm m}$.
The dashed square in (c) shows an example of a position at which the
phase jumps.
}
\label{f:soliton}
\end{figure}
Figure~\ref{f:soliton} shows the dynamics for $N_1 / (N_1 + N_2) = 0.2$.
Since the radius of the initial cylinder of component 1 is larger than
that in Fig.~\ref{f:trap}, the modulation wavelength for the capillary
instability is larger.
The size of each droplet produced by the capillary instability is thus
larger than that in Fig.~\ref{f:trap} and the droplets cut component 2
along the $z$ axis [Fig.~\ref{f:soliton} (c)].
The two-component structure in Fig.~\ref{f:soliton} (c) is similar to
dark--bright solitons~\cite{Busch,Becker,Hamner}, where the density dips
in component 2 and the droplets of component 1 correspond to dark and
bright solitons, respectively.
In fact, the phase $\phi_2$ of component 2 jumps at the density dips
[dashed square in Fig.~\ref{f:soliton} (c)], which is required to
stabilize dark--bright solitons.

\begin{figure}[t]
\includegraphics[width=8.5cm]{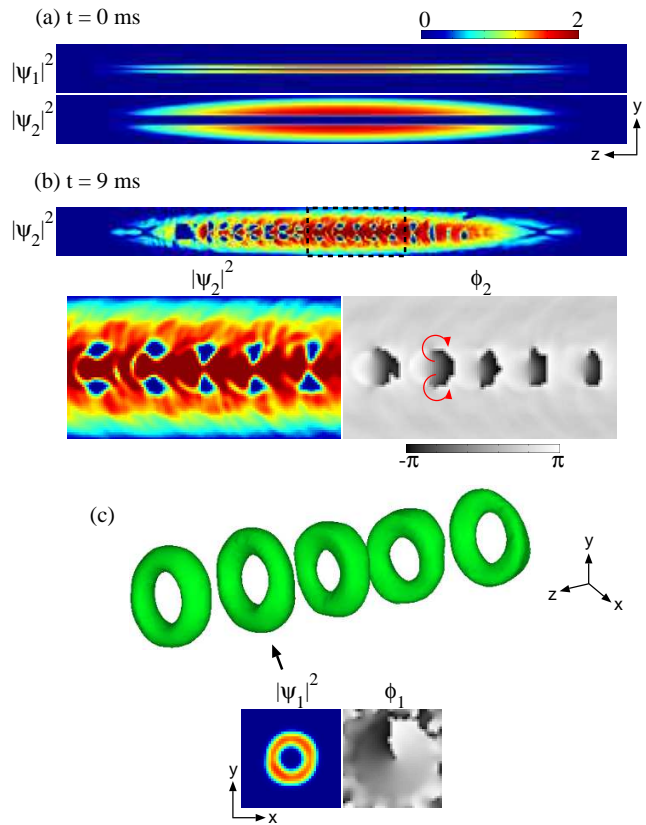}
\caption{
(Color online) Creation of a skyrmion train.
(a) Density distribution of the initial state at the cross section of $x =
0$.
Component 1 has a singly quantized vortex along the $z$ axis.
The initial trap frequencies for component 1 are $(\omega_{1 \perp},
\omega_{1 z}) / 2\pi = (796, 14.5)$ Hz.
At $t = 0$, $\omega_{1 \perp}$ is changed to 145 Hz and the trap center
for component 1 is shifted to $z = 65$ $\mu{\rm m}$.
The other parameters are the same as those in Fig.~\ref{f:trap}.
The field of view is $150 \times 14$ $\mu{\rm m}$.
(b) Density profile of component 2 at the cross section of $x = 0$ at
$t = 9$ ms.
The density and phase in the dashed square are magnified in the lower
panels.
The arrows indicate the directions of circulations of the quantized
vortices.
(c) Isodensity surface of component 1 at $t = 9$ ms.
The lower panels show the density and phase profiles of component 1 at the
cross section of $z$ indicated by the arrow.
The unit of the density in (a)--(c) is $10^{20}$ ${\rm m}^{-3}$.
}
\label{f:sky}
\end{figure}
Figure~\ref{f:sky} demonstrates generation of a skyrmion train using the
capillary instability.
The initial state is a stationary state in which component 1 has a singly
quantized vortex line along the $z$ axis [Fig.~\ref{f:sky} (a)].
Such a state is generated by phase imprinting using, for example, a
stimulated Raman transition with Laguerre--Gaussian
beams~\cite{Andersen}.
At $t = 0$, the radial confinement of component 1 is reduced and a force
in the $z$ direction is exerted on component 1; this can be achieved, for
example, by shifting the trap center in the $z$ direction.
Figures~\ref{f:sky} (b) and \ref{f:sky} (c) show the state at $t = 9$ ms.
The isodensity surface in Fig.~\ref{f:sky} (c) indicates that the initial
hollow cylinder of component 1 breaks up into toruses due to the capillary
instability.
The initial vortex remains in component 1, as shown in Fig.~\ref{f:sky}
(d).
This breakup is not due to the Kelvin--Helmholtz instability since it
occurs in a very similar manner when there is no force in the $z$
direction.
The vortex rings are generated in component 2 [Fig.~\ref{f:sky} (b)] at
locations where there are toruses of component.
This vortex ring generation is due to the acceleration of the toruses of
component 1 in the $z$ direction~\cite{Sasaki2}.
The structures produced in Fig.~\ref{f:sky} are thus the skyrmions in
a two-component BEC~\cite{Ruo}.
However, the skyrmion train in Fig.~\ref{f:sky} is unstable against
axisymmetry breaking.
Because of the vortex quantization, the behavior in Fig.~\ref{f:sky} is
quite different from that in an axially rotating jet of a classical
fluid~\cite{Rutland}.

\section{Conclusions}
\label{s:conc}

We have investigated the capillary instability and resulting dynamics in a
two-component BEC.
We first demonstrated the dynamics in an ideal system in
Sec.~\ref{s:ideal}.
We have shown that modulation occurs on a cylinder of component 1 due to
the capillary instability and the cylinder breaks up into droplets
(Fig.~\ref{f:ideal}), as in a classical fluid jet.
Formation of satellite drops was also observed (Fig.~\ref{f:sate}).
In Sec.~\ref{s:bogo}, we performed Bogoliubov analysis and numerically
obtained a dynamically unstable spectrum, which is in good agreement
with that obtained by Rayleigh's linear stability analysis
(Fig.~\ref{f:bogo}).
In Sec.~\ref{s:trap}, we proposed realistic trapped systems and
numerically demonstrated the dynamics caused by the capillary
instability.
We have shown that the capillary instability can be observed in a
heteronuclear two-component BEC confined in a cigar-shaped harmonic trap,
where the radial confinement of inner component is controlled
(Fig.~\ref{f:trap}).
We have also shown that dark--bright soliton and skyrmion trains can be
generated in this system, which are quantum mechanical objects and have no
counterparts in capillary instability in classical fluids.

\begin{acknowledgments}  
We thank T. Kishimoto and S. Tojo for valuable comments.
This work was supported by the Ministry of Education, Culture, Sports,
Science and Technology of Japan (Grants-in-Aid for Scientific Research,
No.\ 20540388 and No.\ 22340116).
\end{acknowledgments}

\end{document}